# How and Why do Researchers Reference Data? A Study of Rhetorical Features and Functions of Data References in Academic Articles

Sara Lafia (slafia@umich.edu)*[1], Andrea Thomer[3], Elizabeth Moss[1], David Bleckley[1], Libby Hemphill[1,2]

    1. ICPSR, University of Michigan, Ann Arbor, MI, USA
    2. School of Information, University of Michigan, Ann Arbor, MI, USA
    3. School of Information, University of Arizona, Tucson, Arizona, USA
    * Corresponding author

**Authors' Contributions**
Conceptualization, A.T., S.L., E.M., D.B., and L.H.; Formal Analysis, S.L., D.B., and E.M.; Funding Acquisition, L.H., and A.T.; Methodology, S.L., E.M., D.B., A.T., and L.H.; Writing - Original Draft, S.L., L.H, A.T., D.B., and E.M.

**Abstract**
Data reuse is a common practice in the social sciences. While published data play an essential role in the production of social science research, they are not consistently cited, which makes it difficult to assess their full scholarly impact and give credit to the original data producers. Furthermore, it can be challenging to understand researchers' motivations for referencing data. Like references to academic literature, data references perform various rhetorical functions, such as paying homage, signaling disagreement, or drawing comparisons. This paper studies how and why researchers reference social science data in their academic writing. We develop a typology to model relationships between the entities that anchor data references, along with their *features* (access, actions, locations, styles, types) and *functions* (critique, describe, illustrate, interact, legitimize). We illustrate the use of the typology by coding multidisciplinary research articles (n=30) referencing social science data archived at the Inter-university Consortium for Political and Social Research (ICPSR). We show how our typology captures researchers' interactions with data and purposes for referencing data. Our typology provides a systematic way to document and analyze researchers' narratives about data use, extending our ability to give credit to data that support research.

# 1. Introduction

As datasets enter the scientific record, citations connect published data to a larger research network (Hey et al., 2009). Data citations establish precedence for results, provide evidence signaling the quality and significance of research findings, and make it possible to study how researchers use existing data. Citation analysis relies upon the standardization of citations to assess scholarly communication patterns, such as the reach or visibility of ideas across scientific disciplines (e.g., through paper citation networks). Citation indexes of academic papers, like the Science Citation Index (Garfield, 1964), allow researchers to understand who is highly cited, which published work is highly cited, and which publication outlets are prominent. Recent audits of bibliometric networks have also revealed inequalities, suggesting that citations are not objective (Kwon, 2022). For instance, citation rates vary by research topic and author race and gender, suggesting that social factors play an important role in researchers' awareness of published work and decisions to cite it (Kozlowski et al., 2022).

While the infrastructure for studying citation trends for research publications is robust, three main challenges limit the large-scale analysis of data citations. The first challenge is the unambiguous identification of data references. While there are well-established systems for referencing the work of others (Chernin, 1988), many authors still fail to cite data. Many authors refer to data informally in their writing despite data repositories' guidance on best practices for data citation (Fenner et al., 2019) and pressure from funders and publishers to "make data count" (Cousijn et al., 2019). When the burden of linking data to publications falls largely on the author, this often results in partial or inconsistent references to datasets in research articles (Boland et al., 2012). Informal data citation practices make it challenging for readers to understand which data the authors accessed and whether they analyzed data or simply described them (Moss and Lyle, 2018).

A second challenge involves understanding the intent of a data citation. Bibliometric analysis often treats citations as something that can be standardized and universally interpreted as conferring legitimacy to published work (Cronin, 1981). However, like citations of academic literature, researchers cite data for different purposes. Existing citation typologies account for the variety of reasons that researchers cite materials (e.g., to persuade, to critique, to contrast). Many researchers communicate their findings through empirical studies in which they make claims tied to other scholarly products, including published data. Authors' claims range in specificity from explicit to implicit and are often supported by data (Blake, 2010).

A third related challenge involves inferring the quality of citations. Bibliometric measures for quantitative impact assessment, such as the h-index, indicate the popularity or visibility of a source (Egghe, 2010) but say little about the nature of engagement surrounding it. Prior studies of citations to academic literature distinguish surface citations from those that engage deeply with the source material (Cronin, 1984; Leydesdorff, 1998; Spiegel-Rosing, 1977; White and Wang, 1997) and help determine the purpose or polarity of citations (Abu-Jbara et al., 2013; Cohan et al., 2019; Hernández-Alvarez and Gomez, 2016; Teufel et al., 2006). For example, citations that pay homage (e.g., to one's mentors or other influential researchers) also create cumulative advantages, where the best-known researchers receive far more credit for their work (Merton, 1968). Thus, the number of citations a source has received does not indicate the purpose of the citations and may not be a reliable proxy for research quality (Garfield, 1979).

Given the challenges associated with analyzing data references, this study takes a qualitative approach to identify the types of, reasons for, and interactions involving social science data reuse in scientific research. Prior studies of data reference have focused on formal bibliographic citation (Belter, 2014; Jiao and Darch, 2020; Mooney and Newton, 2012; Park et al., 2018). By contrast, we closely analyze even oblique data *mentions* in papers – sentences in which a dataset, or part of a dataset, is named but not formally cited. We find that data references perform a limited set of functions, which we define in a typology that captures and describes the variety of ways researchers refer to data. We then apply the typology to analyze the use of research data in social science publications.

# 2. Background

## 2.1 Defining and citing data

The terms "data" and "datasets" have various meanings depending on their context. Often, what becomes "data" is determined by scientists' choices as they interact with and record observations. "Data," then, are a byproduct of interpretation and can be practically understood as "referring only to that which is analyzed" (Coombs, 1964). Part of this challenge in defining "data" relates to their "unruly" and "poorly bounded" identities, which makes it difficult for them to function as digital objects that can be readily referenced and retrieved (Wynholds, 2011). There is also disagreement on the use of the term "dataset" in technical and scientific literature, which presents challenges for data sharing and preservation (Renear et al., 2010). "Data" are abstract arrangements of symbols that express content; "datasets" are made up of multiple data-bearing entities and may contain additional contextual information about data, such as collection methods (Furner, 2016; Wickett et al., 2012). In our analysis, we focus on archived social science datasets that include contextual metadata and documentation, which have been produced and shared by others for research purposes.

Capturing the relationships between datasets and other scholarly works is critical for giving credit to datasets. Data "references", "mentions", and "citations" signal importance and enable credit through attribution (Altman M, Borgman C, Crosas M, Matone M, 2015). While the terms "references" and "citations" are often used interchangeably, "references" generally indicate that the work is listed in the reference section of a publication (Gilbert and Woolgar, 1974). The term "citation" implies the use of a persistent identifier (PID), which carries a more formal connotation in bibliometrics than "references" or "mentions" (Ball and Duke, 2015). Citations with PIDs link published works to their usage contexts, enabling the verification and reuse of existing scientific analyses (Buneman et al., 2022; Shotton, 2010). While authors are beginning to use digital object identifiers (DOIs) to reference data, best practices for when and why they should do so are not widely followed (Mayo et al., 2016). For example, a recent review of literature citing datasets using their DOIs found that many authors cited data that they had mentioned or described (e.g., data collection methods) but had not been re-analyzed or used in other ways (Banaeefar et al., 2022). In our work, we use the general term "data reference" to cover informal data mentions (i.e., use of a dataset name only), formal citations (i.e., the inclusion of an APA-style citation for a dataset DOI), and descriptions of data use.

## 2.2 Citations in scholarly communication

Citation analysis within scientific disciplines reveals information flows and brings together separate strands of information to construct "consensus models" of subjects within science (Garvey and Griffith, 1972). Citations reflect influences on authors, and citation patterns trace communication across active research networks (Edge, 1979). Citations function differently at the micro and macro levels. At the micro level, citations indicate professional relations and function as rewards, while at the macro level, groups of citations function as concept symbols that codify knowledge in hierarchical social networks (Leydesdorff, 1998). When cited, papers can be invoked as symbolic of the ideas expressed in their text (Small, 1978). Citations are powerful in that they are persistent and take on a separate identity from the people involved in their creation. They are "speech acts," which are brief statements that endure in documents and can be inspected over time (Smith, 2014). Cited sources often substantiate statements or assumptions, point to further information, acknowledge previous research in the same area, and draw critical comparisons indicating the quality of the research (Spiegel-Rosing, 1977).

Studies of research infrastructure rely on citations as metrics for tracing attribution and indicating the impact of scholarly works like datasets or software (Mayernik et al., 2017). Institutions, like journals and data publishers, enforce disciplinary and cultural norms for writing style and citation through publishing guidelines and style manuals. Data citations that use specific identifiers allow readers to identify, retrieve, and give credit to research data. Despite recommendations and best practices, formal data citation is still not commonplace in scholarly writing (Mooney and Newton, 2012). Incomplete, informal, or improperly formatted citations present obstacles to tracking data use (Zhao et al., 2018). Vague or implicit references, for example, make it difficult for readers without an intimate knowledge of variables or other data features to understand which data the authors used and how they used them (Moss and Lyle, 2018). Thus, focusing exclusively on formal citation practices (e.g., using DOIs) means overlooking many potential data references. We seek a more comprehensive understanding of what authors do rhetorically when they formally and informally refer to data in their papers.

## 2.3 Meaning and motivations for citation

Citations bestow credit and recognition in science (Cronin, 1984). However, there may be a disconnect between authors' citation practices and the use of citations to evaluate performance and measure research impact. In other words, citations indicate "what" is cited and how often but do not explain "why" works are cited. Authors' motivations for citing publications can be classified as scientific or tactical. Scientific citations provide background, identify gaps, and establish bases for comparison. Tactical citations acknowledge subjective norms and advertise published work (Lyu et al., 2021). While it is often assumed that citations indicate high-quality work that has influenced authors' research, a survey found that authors' citation decisions were more often motivated by strategic factors rather than their familiarity with the research or perceptions of quality (Teplitskiy et al., 2018).

Behavioral surveys and interviews with authors reveal their judgments about what they are citing and why, which are not reflected in the scholarly record (Liu, 1993). In one such study, authors considered the recency of publications, their topical specificity, and ease of use when deciding whether to cite them (White and Wang, 1997). Authors also believe that citations reflect the prominence or novelty of a document as a "concept marker" and that citing the "concept marker" will bolster the authority of one's work, either through alignment or by critiquing existing work (Case and Higgins, 2000). Silvello identified six main motivations for data citations that are shared across scientific fields: data attribution (accountability and merit), data connection (to claims in publications), data discovery (identification and retrieval), data sharing (reputational), data impact (assessing exposure), and reproducibility (validation and procedures) (Silvello, 2018). These motivations alone, however, do not explain authors' data-citing behaviors and why they vary across venues and contexts.

## 2.4 Analyzing citation content and context

Many computational approaches for citation analysis have been proposed, building on prior insights about authors' motivations to cite. Common citation categories identified across multiple content analysis studies included background information, theoretical framework, prior empirical or experimental evidence, negative distinction, and explanation of methodology (Ding et al., 2014). Features, such as the sections of publications in which citations appear, can be used along with the semantic content of citations to predict citation intent (Nakov et al., 2004). Various classification schemes have been proposed for labeling authors' intents in citing published research (Hernández-Alvarez and Gomez, 2016). One such scheme accounts for citation purposes (i.e., author intent) and polarity (i.e., author sentiment) by distinguishing and weighting negative, neutral, and positive citations (Abu-Jbara et al., 2013). More granular, rule-based coding schemes differentiate statements of weakness, contrasts or comparisons with other work, agreement, compatibility with other work, and neutral citations (Teufel et al., 2006). Conversely, less granular schemes support general citation intent classification by distinguishing background information, method, and comparison citations (Cohan et al., 2019). Such schemes can also help distinguish citation framing (e.g., uses, motivation, future, extends, compare or contrast, background) (Jurgens et al., 2018).

Qualitative approaches respond to the difficulty of predicting authors' intents by focusing on the context and features of references. A review of data citations in academic literature found that they varied along two major dimensions: cited entities and styles (Fear, 2013). Data producers (the researchers who created the data) and data providers (the people or the institution from which the data were obtained) were often named in data citations. Another recent study found that researchers tended to use data created by others for comparison (e.g., ground-truthing, calibration, and identifying baseline measures) and integration (e.g., to ask new questions and conduct new studies) (Pasquetto et al., 2019). A large survey found that existing data are often used as the basis for a new study, to prepare for a new project, to generate new ideas, to develop new methods, to verify their data through analysis and sensemaking, and for teaching (Gregory et al., 2020). Given that data reuse fulfills diverse needs, researchers' purposes for citing data also vary. We build upon established insights into data reuse practices to show how researchers give credit to data.

# 3. Materials and Methods

## 3.1 Sampling frame

We analyzed a sample of publications referencing one or more datasets available through the Inter-university Consortium for Political and Social Research (ICPSR), a large social science data archive at the University of Michigan. We based our typology on existing citation schemes for academic publications, which we extended and refined through iterative coding. The typology captures structural features and rhetorical functions that authors employ when referencing research data. Prior studies have focused on particular publication styles, such as data papers (Jiao and Darch, 2020; Li and Jiao, 2022), or publication outlets, such as PLoS One (Zhao et al., 2018). Instead, we drew from multi-disciplinary publications that referenced social science data archived at ICPSR. This approach allowed us to capture a wider variety of data reference contexts. We considered data references as they occurred in the full-length context of research publications. Further, our only selection requirement was that each publication mentioned one or more archived social science datasets.

We analyzed papers retrieved as part of ICPSR's collection efforts to expand the ICPSR Bibliography of Data-related Literature. The Bibliography includes more than 100,000 publications that use existing social science data available through ICPSR. The review process for the Bibliography involves searching bibliographic databases for references to data available through published ICPSR studies. Staff manually review the metadata and full text of publication search results for evidence of data use. ICPSR maintains strict collection criteria to ensure that publications in the Bibliography reflect data use. Publications are collected if they unambiguously refer to one or more studies available through ICPSR and if it is clear that the authors have accessed and analyzed the data. Publications are rejected from the Bibliography if they fail to demonstrate substantial use of ICPSR data or if the specific studies or series used in the authors' analysis cannot be determined.

To develop a sampling frame for testing our typology, we first identified five publications from the current ICPSR Bibliography representing the multidisciplinary use of ICPSR data. We closely read these publications to identify data references and develop a provisional typology. We then searched an external index of publication full text provided by the Dimensions bibliometric database (Hook et al., 2018) for additional references to any of ICPSR's 11,639 study DOIs available as of February 2022. With the support of ICPSR Bibliography staff, we evaluated and classified the 2,546 search results into six categories indicating whether the publications met the collection criteria for the ICPSR Bibliography. These categories were proposed by ICPSR staff (Banaeefar et al., 2022) and are defined in **Appendix A (Supplementary File 1)**. We then randomly selected publications across each category to include in our analysis, resulting in a total of thirty publications. We gathered additional metadata for each publication, such as the field of research categories from Dimensions, to determine the disciplinary coverage of our sample. We report the publication sampling frame and selection criteria in **Appendix A (Supplementary File 1)**.

## 3.2 Qualitative coding

Our team conducted two phases of coding. The purpose of the first phase was to develop a codebook to describe the diversity of data references. To develop and refine our codes, three annotators from our team read the full-text multiple times for each publication labeled "phase I" listed in **Appendix A (Supplementary File 1)**. The annotators independently proposed refinements to a shared version of the codebook. To achieve qualitative reliability, the annotators drew from many examples and discussed them in weekly meetings (Bauer, 2000). The annotators met weekly to review and incorporate proposed changes in interactive coding sessions. Emerging ideas and conflicting opinions created a dialog from which the codebook was created. In addition to codes, annotators also discussed the scope of the data references and the codebook's focus, purpose, and definitions. Each team member independently applied the updated codes for every iteration of the codebook to identify and annotate all data references in the full text. The team repeated this process until saturation was reached and no new codes were proposed (Charmaz, 2006). The first coding phase resulted in a stable codebook, which we report in **Appendix B (Supplementary File 2)**.

We also described the extent to which annotators' coding aligned. Given that the annotators selected and coded segments from the unstructured, full text of publications, we used Holsti's Index (Holsti, 1969) as an agreement measure. The final Holsti Index was 67.9%, indicating a relatively high level of agreement, given that annotators selected different text segments, which they coded with multiple codes. A single team member independently applied the typology to the held-out set of twenty-five publications labeled "phase II" in the sampling frame reported in **Appendix A (Supplementary File 1)**. This second phase demonstrated the typology in action and captured findings shared in **Section 4**.

# 4. Results

The data reference typology consists of four parent codes (*Data Entity*, *Data Reference*, *Feature*, and *Function*), which are summarized and defined in **Table 1**. A *Data Entity* anchors a *Data Reference* and is based on the pragmatic distinctions raised in work by Renear et al. (Renear et al., 2010) to define components of datasets in scientific literature. Renear et al. distinguish between data content (including files and observations), groupings (the set or study to which they belong), and purposes (metadata used to interpret the data). Similarly, we define *Data Entities* as "one or more words indicating recorded observations" and record them as "Files", "Metadata", "Studies", or "Variables". We excluded *Data Entities* that were not specific or were not discussed in the body of the paper. For example, if the name of a dataset appeared in the title of a publication but data were not described in the main text, we did not consider this a *Data Entity*. Given that some data analysis discussions were broad (e.g., results of statistical tests), we only considered statements that referenced a specific *Data Entity*.

A *Data Reference* is the context window in which one or more *Data Entities* appear. We experimented with various context windows and determined that paragraphs captured sufficient detail leading up to and following a *Data Entity*. We focused on three types of data references, which are introduced in **Section 2.1**: *citations*, *mentions*, and *uses*. *Data citations* include a clear pointer to a published data source but do not name a dataset or indicate that authors used the data. *Data mentions* name a dataset in order to describe it but do not indicate data use. *Data uses* name a dataset and describe interactions between the author and the data. We applied feature and function codes to each *Data Reference*. The full codebook, along with definitions, rules, and examples are provided in **Appendix B (Supplementary File 2)**.

**Table 1**. Overview of parent codes, subcodes, and definitions

| Parent Code | Subcodes | | Definition |
|---|---|---|---|
| | **1st level** | **2nd level** | |
| *Data Entity* | File | | One or more words indicating recorded observations |
| | Metadata | | |
| | Study | | |
| | Variable | | |
| *Data Reference* | | | Context window in which one or more *Data Entities* are mentioned |
| *Feature* | Access | Provision, Reception | Structure, form, and appearance of the data reference |
| | Action | Cites, Mentions, Uses | |
| | Location | Abstract, Acknowledgements, Appendix, Caption… | |
| | Style | Acronym, Generic, Name, Parenthetical | |
| | Type | Derived, Primary, Secondary | |
| *Function* | Critique | Comparison, Limitations | The purposes of the data reference |
| | Describe | Composition, Source | |
| | Illustrate | Context, Outlook | |
| | Interact | Interpretation, Manipulation | |
| | Legitimize | Justification, Transparency | |

## 4.1 Features of data references

Features describe the structure, form, and appearance of the *Data Reference*. The twenty-four features of data references that we identified are organized under five subcodes (*Access*, *Action*, *Location*, *Style*, and *Type*). *Access* codes indicate whether the author describes data sharing or retrieval in the reference. *Action* codes capture the distance between the author and the data along a continuum covering "citing" (i.e., parenthetically referencing data without further context), "mentioning" (i.e., describing or alluding to data), and "using" (i.e., describing active, hands-on with data). The *Action* codes build on the distinction proposed by Pasquetto et al. (Pasquetto et al., 2019) between comparative and integrative data reuse. The *Location* code notes the section of the publication in which the *Data Reference* occurs, such as the abstract, acknowledgments, captions, figures, tables, footnotes, or methods sections stated in the expanded IMRAD structure (Sollaci and Pereira, 2004). The *Style* code captures how the author specifies data entities through the use of an acronym such as "ANES", a generic noun such as "data" or "study", a formal name such as the "American National Election Study, 2016", or a parenthetical citation using author and year of publication. The *Type* code captures whether the data entity is derived from existing data, represents a primary source created by the authors, or is a secondary source published for other researchers to use.

## 4.2 Functions of data references

Functions reflect the purposes of each *Data Reference*. The ten rhetorical functions of data references we identified are organized into five subcodes (*Critique*, *Describe*, *Illustrate*, *Interact*, and *Legitimize*). Definitions and examples for each code and subcode are provided in **Appendix B (Supplementary File 2)**.

The *Critique* code includes (1) *Comparison*, which contrasts the author's work with other work that uses the data or findings from other sources. Authors issue *Comparisons* to draw a contrast between their work and prior findings or to summarize conclusions drawn from the prior use of data. The *Critique* code also includes (2) *Limitations*, which signal authors' awareness and caution when working with data. This code includes acknowledging quality issues, such as potential errors or sampling biases that should limit how data are used.

The *Describe* code includes (3) *Composition*, which explains or discusses knowledge about the data or metadata. *Composition* data references describe what is in the data (e.g., the sampled population) or the study's context (e.g., the data collection method). The *Describe* code also includes (4) *Source*, in which authors describe the provenance of the data. *Source* references acknowledge the origin of the data associated with the data producer or provider.

References labeled with the *Illustrate* code are persuasive. This code includes (5) *Context*, in which the author provides background, findings, or statistics derived from referenced data. In a *Context* reference, the data are metonymic, standing in for the point or claim that the authors are making. The *Illustrate* code also includes (6) *Outlook*, in which authors speculate on potential applications of data that they did not conduct or review in their work. *Outlook* references claim the potential utility of data based on their properties.

The *Interact* code describes hands-on work with data. *Interact* includes a subcode for (7) *Interpretation*, where authors make an empirical claim derived from the analysis of referenced data. *Interpretation* references often follow the description of the authors' analysis. The *Interact* code also includes a subcode for (8) *Manipulation*, where authors describe steps performed while working with data. *Manipulation* involves selecting variables, preparing or transforming data for analysis, and specific data preparation techniques like sampling, correlating, integrating, and validating analyses.

Finally, the *Legitimize* code is used for references intended to persuade the reader through value statements made about data. The *Legitimize* code includes (9) *Justification*, which draws attention to a feature of data that lends credibility or authority to the authors' choices. Examples of *Justification* references include authors' reasons for why data were selected, discussions about the credibility or representativeness of data, and descriptions of previous data uses that qualify their selection. The *Legitimize* code also includes a subcode for (10) *Transparency*, in which authors explain why or how an analysis procedure was applied and signal quality or considerations taken in the analysis. Examples of *Transparency* references include making methods open or reproducible by including analysis in supplementary materials.

## 4.2.1 Data references provide readers with access to data

Data references provide multiple ways of accessing data. Though some journals require that authors include data *provision* statements, where authors make the data used in their analyses available to readers, they were not common in the publications we reviewed. Examples that we encountered included cases where authors provided access to data derived from their analysis for the stated purpose of replication. Alternatively, authors may also provide access to data as a means of recruiting future collaborations. One such description of data *provision* read:

> "The authors have made available the <u>data</u> that underlie the analyses presented in this article (see Styck, Beaujean, & Watkins, 2019), thus allowing replication and potential extensions of this work by qualified researchers. Next users are obligated to involve the data originators in their publication plans, if the originators so desire." (Styck et al., 2019)

Statements about authors' access, or *reception*, of data from providers were often accompanied by formal, parenthetical data references. We defined data *reception* as a reference to an existing data entity and specifications for how that data could be accessed. For example, if a reference is parenthetical, the instance in the reference list must provide an access mechanism, such as a URL, by which others may access the source. In the following example, the author formally attributes the data creator through a parenthetical citation, which includes details about the analysis performed and the historical context motivating selection of the dataset:

> "In order to test whether or not fallout from nuclear testing had persistent effects on the agricultural sector, I create a panel of comparable variables from <u>Historical U.S. Agricultural Censuses</u> for the years 1940 to 1997 Haines et al. (2015). This Census data comes from the most comprehensive surveys of agriculture in the United States that ranges back to 1840. Starting in 1920, the <u>Agricultural Census</u> started conducting bidecennial surveys. I use this data to explore the effects on radioactive fallout deposition on long run outcomes and agricultural development at a national level." (Meyers, 2019)

## 4.2.2 Data references indicate authors' interactions with data

Data references spanned three levels of interactions between authors and data. First, we identified examples of superficial data *citations*, where authors' cited published datasets in the same way as academic articles. In these cases, authors did not name a specific dataset in their writing; instead, they used footnotes or parenthetical citations to formally acknowledge the dataset in their reference list. Most data citations were found in introductory sections and were contextual, meant to provide background, findings, or statistics, which authors used to substantiate a point. It was often unclear, however, how statistics or figures that the authors cited were connected to or derived from the source data. In the following example, the data citation provides findings without direct analysis. No verbs have been used to describe actions performed with or to data; instead, the reader may assume that the authors have some previous experience analyzing the data or that the cited figure is tied to the dataset's published summary statistics. In the following example, the author provides statistics with a corresponding footnote, which leads to a formal citation for data from the India Human Development Survey in the article's reference list:

> "Slums are associated with poor quality housing, water, sanitation, and other services, leading to, among other outcomes, higher rates of disease and death.[17,18] Rich households, on the other hand, are often located in areas with piped water and during water shortages can build storage facilities, tap into underground wells, and pay for delivered water. Only 38% of households among the poorest fifth of India's urban population have access to indoor piped water compared with 62% of the richest fifth." (Frumkin et al., 2020)

When authors *mentioned* data, instead of *citing* them, they described the composition or source of a dataset. Unlike citations, *mentions* name the data in-line. We identified mentions of data primarily in the articles' Methods, Introduction, Discussion, and footnotes. Many data *mentions* provided details about the composition of the data product and relayed knowledge about the basis for the study, collection method, or population. Mentioning data provided background information about data that the authors used later in their analysis or acknowledged the authors' awareness of data that they evaluated but decided not to use. In the following example, the authors describe changes made to the sampled population between waves of a survey in order to qualify their selection method, signal awareness of data quality, and justify their approach:

> "As regards education, health, relationship status, and employment status, Wave 1 respondents who did not remain within the analytical sample show disadvantages compared with those who did. Accordingly, if those more susceptible to depressive symptoms had lower likelihoods of remaining within the analytical sample, attrition between Waves 1 and 2 might lead to conservative assessments of how contexts undergoing economic declines affect their residents' depressive symptoms." (Settels, 2021)

## 4.2.3 Data references are building blocks for empirical arguments

In examples where data were *critiqued*, authors described others' prior efforts or findings to contrast with their approaches. In some cases, authors described how they used the same data differently or decided against using the data based on the reasons that they provided. An example of a data comparison is provided below. The authors present several longitudinal studies covering a similar population and explain potential differences in findings based on differences in their compositions. In this way, the authors signal that they have performed due diligence; they are aware of related studies and can describe their limitations:

> "Studying an earlier cohort than After the JD, the National Longitudinal Bar Passage Study found that long-term bar passage rates were substantially lower for minorities than for whites.[5] Thus a study of all law degree holders including those who did not pass a bar examination may find larger racial gaps in earnings. <u>Census surveys</u> such as those used in this paper lack bar passage status, and therefore likely include a larger proportion of lower earning individuals compared to <u>After the JD</u>.[6]" (McIntyre and Simkovic, 2018)

More references indicating data *use* were found in Methods and Discussion sections of articles as well as in captions, figures, and tables. Mentions and data use statements were distinguished based on the authors' use of verbs and personal pronouns. Most of the use statements described actions, specifically data manipulation (e.g., steps performed while working with data) and interpretation (e.g., making an empirical claim derived from data analysis). Data references describing *use* also occurred in appendixes and supplementary materials rather than in designated areas of articles, like acknowledgments or data availability statements. Examples of data manipulation included selecting variables from referenced data and preparing, transforming, modifying, sampling, subsetting, comparing, or correlating referenced data. Data interpretation included building theories, comparing, and interpreting empirical evidence in figures. The following example illustrates how an author refers to two waves of a study, and related variables, in detailing their analytical approach:

> "To assess the degree to which genetic and environmental factors are stable over time requires an extension of the classical twin design to encompass repeated measurements. Here, we used the bivariate Cholesky decomposition approach (see Figure 1): for each of n measured <u>variables</u>, the Cholesky decomposition specifies n latent A, C, and E factors. Viewed as a diagram, with the latent factors arranged above the measured <u>variables</u>, each of these factors is connected to the measured (manifest) <u>variable</u> beneath it, and to all <u>variables</u> to the right. In this way, each latent factor is connected to one fewer <u>variables</u> than the preceding factor. This design is of value for answering the current question as it allows estimation both of A, C, and E effects at <u>Wave 1</u>, and the extent to which these can account for <u>Wave 2</u> variance, as well the new variance that emerges at <u>Wave 2</u>." (Lewis and Bates, 2017)

# 5. Discussion

Our typology expands the notion of data use beyond re-analysis. For example, while some researchers may access and re-analyze published survey data, many more may reuse that survey's questionnaire or sampling design as a gold standard. Users may also critique the survey data by pointing to its limitations in addressing a particular topic. Our approach casts a wide net to capture these kinds of data references, providing insights into how social science data support research. Our typology is useful for informing recommendation scenarios for researchers about when, why, and how they should reference published data. It also provides a basis for novel data reuse metrics that reflect many forms of engagement with data, from the reuse of survey designs to the re-analysis of survey data.

Our typology also reveals some ways in which data references differ from and align with traditional bibliographic citations. First, the referenced entity can vary in scale; we found references to individual files, metadata records, studies overall, and individual variables. While bibliographic citations may similarly range in scale (e.g., a citation of a specific phrase or section of a paper vs. the paper overall), data entities have a different and possibly broader range of constituent parts. Further expansion and refinement of the typology through review of papers in other domains may reveal additional sub-entities (for instance, research in archaeology or paleontology likely refer to specific artifacts, as well as data derived from those artifacts). Further work is needed to understand the implications of these differing citation scales; are different scales (e.g., variable-level versus full dataset level) references associated with different types of use and argumentation? Are different scales of data more or less likely to result in a formal citation of the dataset? Data entities may additionally have multiple versions that could be referenced (though we did not see this in our sample); how does this complicate our ability to trace the flow of scholarly influence?

Second, we find that data references can act as "concept symbols"(Leydesdorff, 1998), similarly to bibliographic references. Informal reference to datasets by acronym or name (and without a formal citation) indicates a familiarity with datasets as one sees with canonical works of scholarship. In other words, datasets can be referenced with the same familiarity as a biologist references Darwin or an economist references Locke. Future work to identify these foundational or canonical datasets may help reveal how datasets-as-concept symbols differ from bibliographic references. Datasets may be unique in that they also can have a distinct metonymic function, where a reference to a dataset as a whole can stand in for a reference to a specific part or feature of a dataset (as revealed by our "context" code).

Third, data references show interactions with data entities that aren't typically found with bibliographic entities – namely, the provision and archiving of data. Datasets function as both a resource to be used, and a scholarly product to be cited or made available to others. In the publications we annotated, we found that it was uncommon for authors to provide direct access to the findings that they derived from existing data; more often, authors established credibility and trust by simply describing the data source or data provider that they had accessed. Prior studies of researchers' attitudes toward data sharing and reuse show that researchers are reluctant to provide access to their data because they do not believe that the data would be valuable to others or because hoarding data provides a way to attract future collaborations (Cragin et al., 2010; Pasquetto et al., 2019). Though our sample was not representative, we found early indications that align with this prior work.

Finally, we also found alignment with prior schemes describing authors' motivations for citing literature. The rhetorical functions we identified signal the quality, verifiability, or reproducibility of authors' research findings by allowing readers to discover the data the authors have analyzed (Silvello, 2018). For the most part, the data references we reviewed either provided details about dataset composition or descriptions of data manipulation. In the examples we identified, authors affixed additional context about the analysis they performed to connect a data source to its use. Further, when authors included specific access information for data, this enabled readers to retrieve the same dataset.

## 5.1 Limitations and future work

This study proposes a typology that models how authors reference research data. We developed the typology by closely reading papers from the ICPSR bibliography and adding new categories until we reached saturation. The present analysis is not intended to provide quantitative evidence for specific citation trends. We would need to conduct annotation at a larger scale with additional measures in place to verify the agreement of annotators. In addition, we constructed our sampling frame by selecting papers that were first reviewed and classified by experts (i.e., ICPSR Bibliography staff); the sample is balanced across the categories provided in **Appendix A (Supplementary File 1)**. While this sampling strategy is useful for developing and analyzing the ICPSR Bibliography, future uses of the typology for other purposes may require different selection criteria.

We envision applying our typology to study differences in data references across social science disciplines (e.g., sequences or co-occurrences of data reference strategies as markers of scientific disciplines or analytical methods). A recent study of data citation practices at ICPSR observed unexpected uses of dataset DOIs in published literature, which did not indicate data use (Banaeefar et al., 2022). Our typology can be used to study when and why researchers use dataset DOIs and distinguish references that describe data from those that imply data analysis.

# 6. Conclusion

Although research data are increasingly important in modern scientific analyses, they have not been regarded historically as primary research products. The publication, long-term preservation, and dissemination of research data, along with descriptive metadata, make it possible for others to discover, use, and cite observations collected by other researchers for other purposes. We introduced a typology of data references that characterizes the functions data serve in scientific publications: critical, descriptive, illustrative, interactive, and legitimizing. The typology captures researchers' interactions with (e.g., work or analyses done with data) and judgments about data (e.g., claims about its fitness for use based on what is known about data). Understanding why authors reference research data is essential for giving data producers and providers the scholarly research credit they deserve for facilitating scientific work.

**Author Note**
The ICPSR Bibliography of Data-related Literature is available at https://www.icpsr.umich.edu/web/pages/ICPSR/citations/.

**Acknowledgments**
We thank Elizabeth Yakel, Morgan Wofford, and Lizhou Fan from the University of Michigan School of Information for their comments on earlier drafts.

**Funding Information**
This material is based upon work supported by the National Science Foundation under grant 1930645. This project was made possible in part by the Institute of Museum and Library Services LG-37-19-0134-19.

# Appendix A (Supplementary File 1): Sampling Frame

| Phase | Title | Year | Authors | DOI | Pages | Citations | Selection Criteria | Criteria Definition | Publication Fields of Research | ICPSR Studies Referenced | ICPSR Study Authors |
|---|---|---|---|---|---|---|---|---|---|---|---|
| I | The Political Legacy of American Slavery | 2016 | Avidit Acharya, Matthew Blackwell, Maya Sen | 10.1086/686631 | 20 | 187 | Already in Bibliography | Demonstrated use of ICPSR data citing DOI. Title was already in the ICPSR Bibliography (duplicate record). | 16 Studies in Human Society; 1606 Political Science | Three Generations Combined, 1965-1997; Youth-Parent Socialization Panel Study, 1965-1997: Four Waves Combined; Historical, Demographic, Economic, and Social Data: The United States, 1790-2002; ANES 1984 -1998 Time Series Study | Elliot, Patrick; Jennings, M. Kent, Markus, Gregory B., Niemi, Richard G., Stoker, Laura; Haines, Michael R.; Miller, Warren E.; University of Michigan. Institute for Social Research. American National Election Studies |
| I | "Momma's Got the Pill": How Anthony Comstock and Griswold v. Connecticut Shaped US Childbearing | 2010 | Martha J. Bailey | 10.1257/aer.100.1.98 | 31 | 85 | Already in Bibliography | Demonstrated use of ICPSR data citing DOI. Title was already in the ICPSR Bibliography (duplicate record). | 14 Economics; 15 Commerce, Management, Tourism and Services | Natality Detail Files, 1968 - 1980 | United States Department of Health and Human Services. National Center for Health Statistics |
| I | Rural Isolation, Small Towns, and the Risk of Intimate Partner Violence | 2022 | Kathryn O. DuBois | 10.1177/0886260520943721 | 23 | 1 | Already in Bibliography | Demonstrated use of ICPSR data citing DOI. Title was already in the ICPSR Bibliography (duplicate record). | 16 Studies in Human Society; 17 Psychology and Cognitive Sciences; 1602 Criminology; 1607 Social Work; 1701 Psychology | National Crime Victimization Survey, Concatenated File, 1992-2015 | United States Department of Justice. Office of Justice Programs. Bureau of Justice Statistics |
| I | Parent and peer social norms and youth's post-secondary attitudes: A latent class analysis | 2018 | Kremer, Kristen P., Vaughn, Michael G., Loux, Travis M. | 10.1016/j.childyouth.2018.08.026 | 13 | 5 | Already in Bibliography | Demonstrated use of ICPSR data citing DOI. Title was already in the ICPSR Bibliography (duplicate record). | 14 Economics; 16 Studies in Human Society; 1402 Applied Economics; 1607 Social Work | High School Longitudinal Study, 2009-2013 [United States] | United States Department of Education. Institute of Education Sciences. National Center for Education Statistics |

| Phase | Title | Year | Authors | DOI | Pages | Citations | Selection Criteria | Criteria Definition | Publication Fields of Research | ICPSR Studies Referenced | ICPSR Study Authors |
|---|---|---|---|---|---|---|---|---|---|---|---|
| I | Validation of an Index for Functionally Important Respiratory Symptoms among Adults in the Nationally Representative Population Assessment of Tobacco and Health Study, 2014–2016 | 2021 | Halenar, Michael J., Sargent, James D., Edwards, Kathryn C., Woloshin, Steven, Schwartz, Lisa, Emond, Jennifer, Tanski, Susanne, Pierce, John P., Taylor, Kristie A., Lauten, Kristin, Goniewicz, Maciej L., Niaura, Raymond, Anic, Gabriella, Chen, Yanling, Callahan-Lyon, Priscilla, Gardner, Lisa D., Thekkudan, Theresa, Borek, Nicolette, Kimmel, Heather L., Cummings, K.M., Hyland, Andrew, Brunette, Mary F. | 10.3390/ijerph18189688 | 6 | 4 | Already in Bibliography | Demonstrated use of ICPSR data citing DOI. Title was already in the ICPSR Bibliography (duplicate record). | 11 Medical and Health Sciences; 1102 Cardiorespiratory Medicine and Haematology; 1117 Public Health and Health Services | Population Assessment of Tobacco and Health (PATH) Study [United States] Restricted-Use Files | United States Department of Health and Human Services. National Institutes of Health. National Institute on Drug Abuse; United States Department of Health and Human Services. Food and Drug Administration. Center for Tobacco Products |
| II | In the Shadow of the Mushroom Cloud: Nuclear Testing, Radioactive Fallout, and Damage to U.S. Agriculture, 1945 to 1970 | 2019 | Keith Meyers | 10.1017/s002205071800075x | 38 | 3 | Already in Bibliography | Demonstrated use of ICPSR data citing DOI. Title was already in the ICPSR Bibliography (duplicate record). | 14 Economics; 1402 Applied Economics | United States Agriculture Data, 1840 - 2012; Historical, Demographic, Economic, and Social Data: The United States, 1790-2002 | Haines, Michael; Fishback, Price; Rhode, Paul; Haines, Michael R.; Inter-university Consortium for Political and Social Research |
| II | Generalizing Treatment Effect Estimates From Sample to Population: A Case Study in the Difficulties of Finding Sufficient Data | 2016 | Elizabeth A. Stuart, Anna Rhodes | 10.1177/0193841x16660663 | 32 | 33 | Already in Bibliography | Demonstrated use of ICPSR data citing DOI. Title was already in the ICPSR Bibliography (duplicate record). | 16 Studies in Human Society; 13 Education | Head Start Impact Study (HSIS), 2002-2006 [United States] | United States Department of Health and Human Services. Administration for Children and Families. Office of Planning, Research and Evaluation |

| Phase | Title | Year | Authors | DOI | Pages | Citations | Selection Criteria | Criteria Definition | Publication Fields of Research | ICPSR Studies Referenced | ICPSR Study Authors |
|---|---|---|---|---|---|---|---|---|---|---|---|
| II | Prevalence and Predictors of Patient-Reported Long-term Mental and Physical Health After Donation in the Adult-to-Adult Living-Donor Liver Transplantation Cohort Study | 2018 | Mary Amanda Dew, Zeeshan Butt, Qian Liu, Mary Ann Simpson, Jarcy Zee, Daniela P. Ladner, Susan Holtzman, Abigail R. Smith, Elizabeth A. Pomfret, Robert M. Merion, Brenda W. Gillespie, Averell H. Sherker, Robert A. Fisher, Kim M. Olthoff, James R. Burton Jr., Norah A. Terrault, Alyson N. Fox, Andrea F. DiMartini | 10.1097/tp.0000000000001942 | 14 | 25 | Already in Bibliography | Demonstrated use of ICPSR data citing DOI. Title was already in the ICPSR Bibliography (duplicate record). | 1117 Public Health and Health Services; 11 Medical and Health Sciences | Collaborative Psychiatric Epidemiology Surveys (CPES), 2001-2003 [United States] | Alegria, Margarita; Jackson, James S. (James Sidney); Kessler, Ronald C.; Takeuchi, David |
| II | Compound Disadvantage between Economic Declines at the City and Neighborhood Levels for Older Americans' Depressive Symptoms | 2021 | Jason Settels | 10.1177/1535684120980992 | 29 | 0 | Already in Bibliography | Demonstrated use of ICPSR data citing DOI. Title was already in the ICPSR Bibliography (duplicate record). | 12 Built Environment and Design; 16 Studies in Human Society; 1604 Human Geography; 1605 Policy and Administration'; 1205 Urban and Regional Planning | National Social Life, Health, and Aging Project (NSHAP): Round 1, [United States], 2005-2006; National Social Life, Health, and Aging Project (NSHAP): Round 2 and Partner Data Collection, [United States], 2010-2011 | Waite, Linda J.; Laumann, Edward O.; Levinson, Wendy S.; Lindau, Stacy Tessler; O'Muircheartaigh, Colm A.; Waite, Linda J.; Cagney, Kathleen A.; Dale, William; Huang, Elbert; Laumann, Edward O.; McClintock, Martha K.; O'Muircheartaigh, Colm A.; Schumm, L. Phillip; Cornwell, Benjamin |
| II | Longitudinally stable, brain-based predictive models explain the relationships of childhood intelligence with socio-demographic, psychological and genetic factors | 2021 | Narun Pornpattananangk ula, Richard Anneyb, Lucy Riglinb, Anita Thaparb | 10.1101/2021.02.21.432130 | 37 | 2 | Collected | Demonstrated use of ICPSR data citing DOI. Publication title was not in the ICPSR Bibliography (new record). | 08 Information and Computing Sciences; 17 Psychology and Cognitive Sciences; 0801 Artificial Intelligence and Image Processing; 1701 Psychology | Uniform Crime Reporting Program Data: County-Level Detailed Arrest and Offense Data, United States, 2010 | United States Department of Justice. Office of Justice Programs. Federal Bureau of Investigation |

| Phase | Title | Year | Authors | DOI | Pages | Citations | Selection Criteria | Criteria Definition | Publication Fields of Research | ICPSR Studies Referenced | ICPSR Study Authors |
|---|---|---|---|---|---|---|---|---|---|---|---|
| II | Perceived racial discrimination, racial resentment, and support for affirmative action and preferential hiring and promotion: a multi-racial analysis | 2021 | Maruice Mangum, Ray Block Jr. | 10.1080/21565503.2021.1892781 | 24 | 1 | Collected | Demonstrated use of ICPSR data citing DOI. Publication title was not in the ICPSR Bibliography (new record). | 21 History and Archaeology; 2103 Historical Studies | National Politics Study, 2008 | Jackson, James S. (James Sidney); Hutchings, Vincent L.; Wong, Cara; Brown, Ronald |
| II | Profile Reliability of Cognitive Ability Subscores in a Referred Sample | 2019 | Kara M. Styck, A. Alexander Beaujean, Marley W. Watkins | 10.1037/arc0000064 | 10 | 5 | Collected | Demonstrated use of ICPSR data citing DOI. Publication title was not in the ICPSR Bibliography (new record). | 17 Psychology and Cognitive Sciences; 1701 Psychology | Profile Reliability of Cognitive Ability Subscores in a Referred Sample; National Survey on Drug Use and Health, 2014 | Styck, Kara M.; Beaujean, A. Alexander; Watkins, Marley W.; United States Department of Health and Human Services. National Institutes of Health. Eunice Kennedy Shriver National Institute of Child Health and Human Development |
| II | The Temporal Stability of In-Group Favoritism Is Mostly Attributable to Genetic Factors | 2017 | Gary J. Lewis, Timothy C. Bates | 10.1177/1948550617699250 | 22 | 6 | Collected | Demonstrated use of ICPSR data citing DOI. Publication title was not in the ICPSR Bibliography (new record). | 17 Psychology and Cognitive Sciences; 1701 Psychology | Midlife in the United States (MIDUS 2), 2004-2006 | Ryff, Carol D.; Almeida, David M.; Ayanian, John Z.; Carr, Deborah S.; Cleary, Paul D.; Coe, Christopher; Davidson, Richard J.; Krueger, Robert F.; Lachman, Marge E.; Marks, Nadine F.; Mroczek, Daniel K.; Seeman, Teresa E.; Seltzer, Marsha Mailick; Singer, Burton H.; Sloan, Richard P.; Tun, Patricia Ann; Weinstein, Maxine; Williams, David R. |

| Phase | Title | Year | Authors | DOI | Pages | Citations | Selection Criteria | Criteria Definition | Publication Fields of Research | ICPSR Studies Referenced | ICPSR Study Authors |
|---|---|---|---|---|---|---|---|---|---|---|---|
| II | Barriers and facilitators to successful transition from long-term residential substance abuse treatment | 2016 | Jennifer I. Manuel, Yeqing Yuan, Daniel B. Herman, Dace S. Svikis, Obie Nichols, Erin Palmer, Sherry Deren | 10.1016/j.jsat.2016.12.001 | 7 | 33 | Data mention | Acknowledged a brief point (e.g., a statistic) with DOI, but data were not analyzed in the publication. | 17 Psychology and Cognitive Sciences; 1701 Psychology | National Survey on Drug Use and Health, 2011 | United States Department of Health and Human Services. Substance Abuse and Mental Health Services Administration. Center for Behavioral Health Statistics and Quality |
| II | Examining Risk for Frequent Cocaine Use: Focus on an African American Treatment Population | 2016 | Tamika Chere Barkley Zapolski, Patrick Baldwin & Carl W. Lejuez | 10.3109/10826084.2016.1155618 | 11 | 6 | Data mention | Acknowledged a brief point (e.g., a statistic) with DOI, but data were not analyzed in the publication. | 11 Medical and Health Sciences; 1117 Public Health and Health Services | National Survey on Drug Use and Health, 2013 | United States Department of Health and Human Services. Substance Abuse and Mental Health Services Administration. Center for Behavioral Health Statistics and Quality |
| II | Anzansi family program: a study protocol for a combination intervention addressing developmental and health outcomes for adolescent girls at risk of unaccompanied migration | 2020 | Ozge Sensoy Bahar , Fred M. Ssewamala, Abdallah Ibrahim, Alice Boateng, Proscovia Nabunya, Torsten B. Neilands, Emmanuel Asampong, Mary M. McKay | 10.1186/s40814-020-00737-4 | 12 | 2 | Data mention | Acknowledged a brief point (e.g., a statistic) with DOI, but data were not analyzed in the publication. | 17 Psychology and Cognitive Sciences; 11 Medical and Health Sciences; 1701 Psychology; 1117 Public Health and Health Services | Youth Development Study, 1988-2011 [St. Paul, Minnesota] | Mortimer, Jeylan T. |
| II | Protecting Adolescents in Low- And Middle-Income Countries from Interpersonal Violence (PRO YOUTH TRIAL): Study Protocol for a Cluster Randomized Controlled Trial of the Strengthening Families Programme 10-14 ("Familias Fuertes") in Panama | 2018 | Anilena Mejia, Richard Emsley, Eleonora Fichera, Wadih Maalouf, Jeremy Segrott, Rachel Calam | 10.1186/s13063-018-2698-0 | 13 | 4 | Instrument mention | Acknowledged a study's instrument or a question from an instrument with a DOI. | 11 Medical and Health Sciences; 1117 Public Health and Health Services | National Youth Survey [United States]:  Wave VII, 1987 | Elliott, Delbert S. |

| Phase | Title | Year | Authors | DOI | Pages | Citations | Selection Criteria | Criteria Definition | Publication Fields of Research | ICPSR Studies Referenced | ICPSR Study Authors |
|---|---|---|---|---|---|---|---|---|---|---|---|
| II | Examining cohort effects in developmental trajectories of substance use | 2017 | Alison Reimuller Burns, Andrea M. Hussong, Jessica M. Solis, Patrick J. Curran, James S. McGinley, Daniel J. Bauer, Laurie Chassin, Robert A. Zucker | 10.1177/0165025416651734 | 11 | 7 | Instrument mention | Acknowledged a study's instrument or a question from an instrument with a DOI. | 17 Psychology and Cognitive Sciences; 1701 Psychology; 1702 Cognitive Sciences | Monitoring the Future: A Continuing Study of the Lifestyles and Values of Youth, 1978 | Bachman, Jerald G.; Johnston, Lloyd D.; O'Malley, Patrick M. |
| II | Daily Interpersonal Tensions and Well-Being Among Older Adults: The Role of Emotion Regulation Strategies | 2020 | Kira S. Birditt, Courtney A. Polenick, Gloria Luong, Susan T. Charles, Karen L. Fingerman | 10.1037/pag0000416 | 30 | 15 | Instrument mention | Acknowledged a study's instrument or a question from an instrument with a DOI. | 17 Psychology and Cognitive Sciences; 1701 Psychology | Family Exchanges Study Wave 1, Philadelphia, Pennsylvania, 2008 | Fingerman, Karen L. |
| II | School Inequalities and Urban Welfare: Going beyond Socioeconomic Status with Data Science | 2019 | Renato P. dos Santos, Şahin Bülbül, Isadora L. Lemes | 10.17648/acta.scientiae.5494 | 26 | 1 | Methodology mention | Used data DOI to acknowledge a study's methodology or design. | 16 Studies in Human Society; 1608 Sociology | Equality of Educational Opportunity (COLEMAN) Study (EEOS), 1966 | Coleman, James S. |
| II | The relationship between unemployment and fertility in Italy: A time-series analysis | 2016 | Alberto Cazzola, Lucia Pasquini, Aurora Angeli | 10.4054/demres.2016.34.1 | 40 | 43 | Methodology mention | Used data DOI to acknowledge a study's methodology or design. | 16 Studies in Human Society; 1603 Demography | Solution to the Ecological Inference Problem: Reconstructing Individual Behavior from Aggregate Data | King, Gary |
| II | Social-ecological Functions and Vulnerability Framework to Analyze Forest Policy Reforms | 2012 | Fanny Rives, Martine Antona, Sigrid Aubert | 10.5751/es-05182-170421 | 18 | 8 | Methodology mention | Used data DOI to acknowledge a study's methodology or design. | 41 Environmental Sciences; 4101 Climate Change Impacts and Adaptation | Agendas, Alternatives, and Public Policies, 1976, 1977, 1978, 1979 [United States] | Kingdon, John W. |
| II | Protecting health in dry cities: considerations for policy makers | 2020 | Howard Frumkin, Maitreyi Bordia Das, Maya Negev, Briony C Rogers, Roberto Bertollini, Carlos Dora, Sonalde Desai | 10.1136/bmj.m2936 | 8 | 5 | Source mention | DOI linked to a source of data from another study or DOI linked to a data source that is suggested but not analyzed. | 11 Medical and Health Sciences; 1117 Public Health and Health Services; 1103 Clinical Sciences | India Human Development Survey Panel (IHDS, IHDS-II), 2005, 2011-2012 | Desai, Sonalde; Vanneman, Reeve; National Council of Applied Economic Research, New Delhi |
| II | Physics of Laser in Contemporary Visual Arts: the research protocol | 2016 | Diaa Ahmed Mohamed Ahmedien | 10.3897/rio.2.e11150 | 11 | 1 | Source mention | DOI linked to a source of data from another study or DOI linked to a data source that is suggested but not analyzed. | 19 Studies in Creative Arts and Writing; 1902 Film, Television and Digital Media | Reporting the Arts II [2003] | Szanto, Andras |

| Phase | Title | Year | Authors | DOI | Pages | Citations | Selection Criteria | Criteria Definition | Publication Fields of Research | ICPSR Studies Referenced | ICPSR Study Authors |
|---|---|---|---|---|---|---|---|---|---|---|---|
| II | Are law degrees as valuable to minorities? | 2017 | Frank McIntyre, Michael Simkovic | 10.1016/j.irle.2017.09.004 | 57 | 3 | Source mention | DOI linked to a source of data from another study or DOI linked to a data source that is suggested but not analyzed. | 18 Law and Legal Studies; 1801 Law | After the JD, Wave 3: A Longitudinal Study of Careers in Transition, 2012-2013, United States | Nelson, Robert; Dinovitzer, Ronit; Plickert, Gabriele; Sterling, Joyce; Garth, Bryant G. |
| II | Distance, Size and Turmoil: North-South Mediterranean Interactions | 2014 | J. Patrick Rhamey, William R. Thompson, Thomas J. Volgy | 10.4000/cdlm.7840 | 21 | 3 | Collected | Demonstrated use of ICPSR data citing DOI. Publication title was not in the ICPSR Bibliography (new record). | 16 Studies in Human Society; 1606 Political Science | Conflict and Peace Data Bank (COPDAB), 1948-1978 | Azar, Edward E. |
| II | Fukushima effects in Germany? Changes in media coverage and public opinion on nuclear power | 2015 | Dorothee Arlt, Jens Wolling | 10.1177/0963662515589276 | 16 | 37 | Data mention | Acknowledged a brief point (e.g., a statistic) with DOI, but data were not analyzed in the publication. | 13 Education; Studies in Creative Arts and Writing; 22 Philosophy and Religious Studies; 1302 Curriculum and Pedagogy; 2202 History and Philosophy of Specific Fields; 1903 Journalism and Professional Writing | Global Snap Poll on Tsunami in Japan and Impact on Views About Nuclear Energy, 2011 | Gallup International, Inc. |
| II | Evaluating the Short-term Impact of Media Aware Parent, a Web-based Program for Parents with the Goal of Adolescent Sexual Health Promotion | 2019 | Tracy M. Scull, Christina V. Malik, Elyse M. Keefe, Alexander Schoemann | 10.1007/s10964-019-01077-0 | 21 | 17 | Instrument mention | Acknowledged a study's instrument or a question from an instrument with a DOI. | 17 Psychology and Cognitive Sciences; 1701 Psychology | Iowa Youth and Families Project, 1989-1992 | Conger, Rand; Lasley, Paul; Lorenz, Frederick O.; Simons, Ronald; Whitbeck, Les B.; Elder Jr., Glen H.; Norem, Rosalie |
| II | Are Self-report Measures Able to Define Individuals as Physically Active or Inactive? | 2015 | Jostein Steene-Johannessen, Sigmund A. Anderssen, Hidde P. van der Ploeg, Ingrid J.M. Hendriksen, Alan E. Donnelly, Søren Brage6, and Ulf Ekelund | 10.1249/mss.0000000000000760 | 35 | 154 | Methodology mention | Used data DOI to acknowledge a study's methodology or design. | 11 Medical and Health Sciences; 1117 Public Health and Health Services | Eurobarometer 64.3: Foreign Languages, Biotechnology, Organized Crime, and Health Items, November-December 2005 | Papacostas, Antonis |

| Phase | Title | Year | Authors | DOI | Pages | Citations | Selection Criteria | Criteria Definition | Publication Fields of Research | ICPSR Studies Referenced | ICPSR Study Authors |
|---|---|---|---|---|---|---|---|---|---|---|---|
| II | Time-varying effects of income on hippocampal volume trajectories in adolescent girls | 2017 | Monica E. Ellwood-Lowea, Kathryn L. Humphreysa, Sarah J. Ordaza, M.Catalina Camachoa, Matthew D. Sacchetb, Ian H. Gotliba | 10.1016/j.dcn.2017.12.005 | 10 | 34 | Source mention | DOI linked to a source of data from another study or DOI linked to a data source that is suggested but not analyzed. | 17 Psychology and Cognitive Sciences; 1701 Psychology | Equality of Educational Opportunity (COLEMAN) Study (EEOS), 1966 | Coleman, James S. |

# Appendix B (Supplementary File 2): Codebook

| Parent Code | 1st Child Code | 2nd Child Code | Definition | Notes | Example |
|---|---|---|---|---|---|
| Data Entity | | | One or more words indicating recorded observations | First identify and code all relevant Data Entities in the paper. Search for the study name and the study author names if given in the sampling frame. One or more words that are the proper noun of a data product (eg, the name of a survey) or a keyword standing in (eg, the "data", "file"). | |
| | File | | Item encoding observations | Search for the term "file" and synonym: "data". Exclude phrases like "data collection" or abstract discussions of data. | *Our analyses utilized the adult W2 and W3 Restricted Use Files (https://doi.org/10.3886/ICPSR36231.v21 (accessed on 17 February 2021)).* |
| | Metadata | | Item describing a collection of observations | Search for the term "metadata" and synonyms: "codebook", "documentation", "user guide". | *PATH Study design and methods [6,16,17], interviewing procedures, questionnaires, sampling, weighting, and response rates are in the PATH Study Restricted Use Files User Guide.* |
| | Study | | Item holding a collection of observations or a parenthetical statement standing in for a collection of observations as (Author, Year) | Search for the term "study" and synonyms: "dataset", "series", "survey", "wave". Exclude references like the "present study" or "case study". | *The Population Assessment of Tobacco and Health (PATH) Study is an ongoing, nationally representative, longitudinal cohort study that obtains detailed information on tobacco use in the U.S. population.* |
| | Variable | | Item representing a dimension of observations | Search for the term "variable" and synonyms: "indicator" and "weight". Exclude phrases referring to variables in an abstract sense like "predictor, moderator, and outcome variables". | *Three wheezing variables (ever wheezing, past 12-month (P12M) wheezing, and P12M wheezing attack frequency) were combined to create one variable (0 = never wheezing; 1 = ever wheezing, but no P12M wheezing OR P12M wheezing but no wheezing attacks; 2 = P12M wheezing and 1–3 attacks; 3 = P12M wheezing and 4–12 attacks; 4 = P12M wheezing and more than 12 attacks).* |
| Data Reference | | | Context window in which one or more Data Entities are mentioned | Code the paragraph containing one or more Data Entities as a Data Reference. Each Data Reference contains one or more Data Entities. | |
| Feature | | | Structure, form, and appearance of the data reference in the text | Assign one or more 2nd child codes if they apply to the Data Reference. | |
| | Access | | Specification of where the data can be accessed | Author must give a URL/DOI or point of contact from which data can be accessed. Organizational information alone is insufficient. | |
| | | Provision | Sharing data back with the research community | Author created a data entity and specifies where it can be accessed by others | *Data and supporting materials necessary to reproduce the numerical results in the paper are available in the JOP Dataverse (https://dataverse.harvard.edu /dataverse/jop).* |

| Parent Code | 1st Child Code | 2nd Child Code | Definition | Notes | Example |
|---|---|---|---|---|---|
| | | Reception | Accessing data from the research community | Author refers to existing data entity and specifies where if can be accessed. If the reference is parenthetical, the instance in the reference list needs to provide an access mechanism. Stating a person or institution is not sufficient if access mechanism is not provided. | *Elliot, Patrick. 2007. "Three Generations Combined, 1965–1997." Inter- university Consortium for Political and Social Research (ICPSR). http://doi.org/10. 3886/ICPSR04532.v1.* |
| | Action | | Interaction between authors and data | Determine the proximity of the Data Reference indicating the authors' relationship to the data product | |
| | | Cites | Points to data superficially but doesn't name or contextualize it. Provides findings without analysis. | Parenthetical citation refers to data but the statement does not mention a data entity. No verb is used to describe an action taken with data. | *An estimated 21.5 million people have a diagnosable substance use disorder (SUD), representing 9% of the U.S. population, and approximately 40,000 more people engage in misuse that is considered medically harmful (McLellan & Woodworth, 2014; Substance Abuse and Mental Health Services Administration, 2015).* |
| | | Mentions | Mentions, describes, or alludes to data | Reference to Data Entity is descriptive | *In many years, the ANES contains "feeling thermometer" questions, which ask respondents to evaluate their feelings about politicians and groups (including racial or ethnic groups) on a scale from 0 to 100.* |
| | | Uses | Indicates active, hands-on data work | Actions with the data are evident. Use of verbs and "we" statements in relation to Data Entity. | *For the 2010 CCES, when both questions were asked, we rescaled both questions and averaged them to create one measure.* |
| | Location | | Where the data reference occurs | Determine IMRAD location of the paper in which the Data Reference appears | |
| | | Abstract | In publication abstract | In the section providing a brief overview of the paper | *Logistic regression estimates of semiannual IPV prevalence were modeled using generalized estimating equations and robust standard errors to compensate for repeated measures and for the complex sample design of the NCVS.* |
| | | Acknowledgements | In acknowledgements section | In an acknowledgements statement or section, which may include additional information about funding, PIs | *We are also grateful to Michael Haines, Eitan Hersh, andHeather O'Connell for sharing data with us.* |
| | | Appendix | In appendix section or supplement | In the section of the paper authors refer to as appendix, supplement, etc. | *1955 Growth of American Families Study. This study sampled 18- to 39-year-old white women who were currently married. For information on whether respondents ever used contraception in the 1955 GAF, information was taken from two questions.* |

| Parent Code | 1st Child Code | 2nd Child Code | Definition | Notes | Example |
|---|---|---|---|---|---|
| | | **Caption, Figure, Table** | In a table or figure caption | In a table, a figure, or their caption. Overrides other section information (eg, code footnote rather than footnote and conclusion). Code the entire table or figure in which the Data Entity is mentioned. | *Note. Inequality here is measured by the log of the ratio of white to black median incomes within counties in 2014. Data come from the American Community Survey, 2009–14.* |
| | | **Conclusion** | In conclusion section | In the section providing a summary or conclusion at the end of the paper | *However, if the divide is based only on the delineation conventionally used to represent urban, suburban, and rural settlements in analyses of the NCVS, then the results of this study suggest that any such urban/rural divide is a myth and that our attention would better spent on understanding the suburban/non-suburban divide in the risk of IPV.* |
| | | **Data Availability** | In data availability statement | In a separate article field. May be mandated by journal and include access information. | *Data Availability Statement: Data from the PATH Study Wave 1 to Wave 3 are available for down- load as Restricted Use Files (https://www.icpsr.umich.edu/icpsrweb/NAHDAP/studies/36231). Request guidelines and conditions of use are available at the website above.* |
| | | **Discussion** | In results or discussion section | In the section describing or interpreting findings | *These limitations were addressed by comparing PATH Study data to that of NHANES and using questions already validated for use in capturing respiratory illness in other populations.* |
| | | **Footnote** | In a footnote | In a footnote or endnote. Overrides other section information (eg, code footnote rather than footnote and conclusion). | *See the IPUMS documentation at https://usa.ipums.org/ for a complete description of these census measures.* |
| | | **Funding** | In funding information | Author provides details about study funder and/or grant details | *The author(s) disclosed receipt of the following financial support for the research, authorship, and/or publication of this article: Supported in part by Award DRL- 1335843 from the National Science Foundation (co-PI's Stuart and Olsen) and by Award R305D150003 from the Institute of Education Sciences, U.S. Department of Education (co-PIs Stuart and Olsen).* |

| Parent Code | 1st Child Code | 2nd Child Code | Definition | Notes | Example |
|---|---|---|---|---|---|
| | | Introduction | In introductory or background sections | In the section providing an introduction or background at the beginning of the paper | *In this article, I will consider data from the National Crime Victimization Survey (NCVS) to estimate variations in the risk of IPV across settlement types using a measure that distinguishes isolated rural settlements from the more populous nonmetropolitan settlements that are conventionally coded as rural despite their relatively high residential densities.* |
| | | Keyword | In keywords | Code whole block of keywords, including Data Entity, as Data Reference | *Keywords: intimate partner violence, prevalence, NCVS, settlement type* |
| | | Methods | In materials and methods, or data, section | In the section describing materials and methods | *The data used in this study are from n = 578,471 females aged 18 and older interviewed a total of n = 1,672,999 times in the NCVS over the period 1994 to 2015 (BJS, 2016).* |
| | Style | | Specification of how the author names the entity/ies in the data reference | | |
| | | Acronym | Uses an acronym to refer to a data entity | Data entity is given as an acronym only | *This section details the use of the REDI and HSIS data to evaluate the generalizability of the REDI program to Head Start–eligible children nationwide, as represented by the HSIS.* |
| | | Generic | Uses a general noun to refer to data entity, such as "data", "study", "wave", "variable" | Data entity is informally specified using lowercase word and without version information | *Data were made available by the Interuniversity Consortium for Political and Social Research, Ann Arbor, MI, USA. Neither the collector of the original data nor the Consortium bears any responsibility for the analyses or interpretations presented here.* |
| | | Name | Uses formal or informal name of a data entity | Data entity uses an unambiguous name May or may not include specific version information (such as which wave was used) | *Studying an earlier cohort than After the JD, the National Longitudinal Bar Passage Study found that long-term bar passage rates were substantially lower for minorities than for whites.* |
| | | Parenthetical | Uses author and year linked to the reference list to refer to data entity Can also be endnote, footnote style pointer, or APA style citation | Data entity is not stated in-line but is referenced in parentheses and listed in works cited | *In order to test whether or not fallout from nuclear testing had persistent effects on the agricultural sector, I create a panel of comparable variables from Historical U.S. Agricultural Censuses for the years 1940 to 1997 Haines et al. (2015).* |
| | Type | | Authors' relation to data | Use context to figure out whether the authors created the data or use data created by others | |

| Parent Code | 1st Child Code | 2nd Child Code | Definition | Notes | Example |
|---|---|---|---|---|---|
| | | Derived | Creates new data from existing data | New data derived from analysis or combinations of existing data, which authors may share Merging or applying functions creates derived data but subsetting does not Merging nested levels of data files (eg: individual to household) or mulitple years of longitudinal data would not be considered deriving | *We discuss our data in the next section and present our core results linking the prevalence of slavery in 1860 and contemporary attitudes in the following section, with additional robustness checks presented in the appendix, available online.* |
| | | Primary | Reflects the creation of data | Data that the authors collected themselves This covers proto-data that authors are interpreting but may not be published or explicitly named | *The authors have made available the data that underlie the analyses presented in this article (see Styck, Beaujean, & Watkins, 2019), thus allowing replication and potential extensions of this work by qualified researchers. Next users are obligated to involve the data originators in their publication plans, if the originators so desire.* |
| | | Secondary | Uses data created by others or made available for others to use (eg, archived data) | Data that the authors attributed to another creator or refer to as archived data This covers instances where authors are reusing their own data but are referring to it as a published product | *For all of these variables, we mapped 1860 data onto modern county boundaries using the procedure described in appendix section A.* |
| Function | | | The purposes of the data reference | Pick one or more instances of each 2nd child code to apply to each Data Reference A Data Reference can have more than one rhetorical function | |
| | Critique | | Critiques the referenced data | Purpose is to persuade the reader through critique of the data product | |
| | | Comparison | Draws a contrast between the author's work and other work done with the data or presents a comparison across data sources | Contrasting with prior findings, conclusions drawn from prior use of referenced data May include terms like "unlike", "but" | *The effect sizes from the 1950s and 1960s are roughly similar in magnitude, but in the opposite direction, to the effect of slavery on the Obama white vote in 2008, estimated from CCES white respondents.* |
| | | Limitations | Signals authors' awareness and caution taken (eg, to bound interpretations) while working with the data | Acknowledging potential errors, shortcomings, quality issues, fitness for use, privacy - context trade-offs, sampling issues of referenced data May include terms like "weakness", "shortcoming" | *First, the 1860 slave data are historical and may be measured with error.* |
| | Describe | | Describes the referenced data neutrally | Purpose is to provide the reader with details to support their understanding | |
| | | Composition | Explains or discusses knowledge about the data product, such as its metadata | Describing what's in the data (the basis for the study, its collection method, population, etc) May use terms like "includes", "representing" | *After restricting the sample to Southern whites, we have an ANES sample of 3,123 individuals across 64 counties in the South.* |
| | | Source | States where the data came from, creating an association back to the origin of the data | Giving credit to data producer/provider, describing its accessibility, and its provenance May use terms like "from", "provided by" | *We constructed these measures using data from the UN Food and Agriculture Organization (FAO).* |
| | Illustrate | | Illustrates a perspective suppported by referenced data | Purpose is to illustrate with empirical evidence | |

| Parent Code | 1st Child Code | 2nd Child Code | Definition | Notes | Example |
|---|---|---|---|---|---|
| | | Context | Background, findings, or statistics drawn from referenced data where data stand in for the author's point (ie, are metonymic) | Data is referenced in the same way as a research article (the whole entity is referenced to make a point) | *For example, there may be interest in the effects of the REDI intervention on all 4-year-olds, including a more socioeconomically diverse population of children than those eligible for Head Start, in which case a nationally representative data set of children in the United States, such as the ECLS-B, would be a more appropriate population data set (http://nces.ed.gov/ecls/birth.asp).* |
| | | Outlook | Comments on potential applications of data that were not conducted or reviewed in the present work | Stating that data "could be useful for" or have properties that would be useful for an application | *Note that the HSIS used an earlier but comparable version of this measure, which could be used in future work to do further diagnostics regarding the generalizability of the Research-Based, Developmentally Informed results to the HSIS.* |
| | Interact | | Describes interactions taken with the referenced data | Purpose is to tell the reader what was done with or derived from the data product | |
| | | Interpretatation | Makes an empirical claim derived from analysis of data | Building theories, comparisons, interpreting empirical evidence from referenced data, follows descriptions of analysis, and/or accompanies figures May include terms like "show", "find" | *Instead, as figure 4 indicates, the data show that regional differences emerged after the end of Reconstruction.* |
| | | Manipulation | Describes steps performed while working with data | Selecting variables, files, waves, instruments from referenced data; preparing, transforming, modifying, sampling, comparing, correlating, validating, integrating, extending, updating analysis of referenced data May include terms "subset", "sample" | *We subset these data to the former Confederate states plus Missouri and Kentucky, both of which had significant internal support for the Confederacy,5 and to self-identified whites, leaving us with more than 40,000 respondents across 1,329 of the 1,435 Southern counties.* |
| | Legitimize | | Confers legitimacy to the work through association with referenced data | Purpose is to persuade the reader through value statements about the data product | |
| | | Justification | Draws attention to a feature of data that lends credibility or authority to authors' choices | Reasoning, choosing data to establish trust, credibility of data; signaling normativity through typical uses, prior studies using data; substantiating or motivating, justifying, qualifying data selection May include terms "although", "representative" | *Since county boundaries have shifted since 1860, we use an area-weighting method to map data from the 1860 Census onto county boundaries in 2000, enabling us to estimate the proportion enslaved in 1860 within modern-day counties* |
| | | Transparency | Explains why or how an analysis procedure was applied, signaling quality or considerations taken in analysis | Making methods open and perhaps reproducible to adhere journal or funder data sharing requirements Includes pointing to data analysis in appendix or supplemental materials May include terms "document", "provide" | *In addition, we also provide additional evidence in appendix D, in which we analyze contemporary mobility data from the 2000 US Census.* |